\documentclass[twocolumn]{aastex701} 
\usepackage{url}   
\usepackage{epstopdf}
\usepackage{graphicx}
\usepackage{lineno}
\hypersetup{linkcolor=magenta,citecolor=blue,filecolor=cyan,urlcolor=purple}

\received{November 17, 2025} 
\revised{****,**, 2025} 
\accepted{****} 

\newcommand{\cm}[1]{{\color{magenta}{#1}}}

\shorttitle{Modeling the magnetic evolution in AR 13500}
\shortauthors{Vemareddy et al}

\begin{document}
\title{Understanding the formation and eruption of sigmoidal structure through data-driven modeling of magnetic evolution in solar active region 13500}

\author[orcid=0000-0003-4433-8823,sname='P']{P. Vemareddy} 
\affiliation{Indian Institute of Astrophysics, II Block, Koramangala, Bengaluru-560 034, India}
\email[show]{vemareddy@iiap.res.in} 
\author[orcid=0009-0008-6709-6648,sname='S']{S. Nair} 
\affiliation{Indian Institute of Astrophysics, II Block, Koramangala, Bengaluru-560 034, India}
\affiliation{National Institute of Technology, Warangal-506 004, India}
\email[show]{sreenair1202@gmail.com} 

\author[orcid=0000-0003-4433-8823,sname='P']{S. Gosain} 
\affiliation{National Solar Observatory, 3665 Discovery Drive, Boulder 80303, Colorado, USA}
\email[show]{sgosain@nso.edu} 

\begin{abstract}
We investigate the magnetic origin of the coronal mass ejection (CME) that occurred on November 28, 2023, at 19:50UT from active region (AR) 13500 located near the solar disk-center. The eruption was associated with an S-shaped sigmoidal structure formed by the inner AR polarities along a sheared polarity inversion line, while the outer polarities evolved through proper motions. During November 26-28, the AR exhibited a decrease in net magnetic flux while progressively injecting magnetic helicity and energy into the corona toward the eruption onset, highlighting the key role of helicity-injection in triggering eruptions. To simulate this magnetic evolution, we employed a data-driven magnetofrictional (MF) simulation starting 2.8 days prior to the eruption. The energy input for the model was constrained using the observed energy injection through an ad-hoc parameter. The initial potential-field configuration gradually evolved into a sheared-arcade and eventually developed into a twisted flux rope (FR) over the observed time-scale. Proxy emission maps based on electric currents show remarkable morphological agreement between the simulated and observed sigmoidal structure. The average FR-core twist increasingly builds-up leading the FR to initiate slow-rise motion of FR top from 50Mm  until its eruption onset at 80Mm. Importantly, the ratio of current-carrying to total relative-helicity increased from 0.13 at FR formation to 0.30 at eruption, when the FR core entered the torus-unstable regime, suggesting an association between torus-instability and a threshold helicity ratio. These results demonstrate that data-driven MF simulations can successfully reproduce the evolving coronal magnetic configuration and may provide a robust tool for assessing the eruptive potential of ARs, particularly the helicity ratio.
\end{abstract}

\section{Introduction}
\label{Intro}
\linenumbers 
Solar flares and coronal mass ejections (CMEs) are massive explosive phenomena representing significant solar activity that varies with the 11-year sunspot cycle. During these events, vast amounts of magnetized plasma are expelled from localized regions in the solar atmosphere
into space, accompanied by bursts of electromagnetic radiation, including X-rays and ultraviolet light. These phenomena are fundamental drivers of space weather, affecting Earth's magnetosphere and technological systems \citep{ZhangJie2021_EarthEff_SolTrans, Temmer2021}. The energy powering these events originates from concentrated magnetic field regions, such as active regions (ARs), though the exact mechanisms of energy storage and 
eruption onset are still debated \citep[e.g.,][]{forbes2000, green18}.

The study of magnetic field evolution is central to understanding how magnetic energy is stored and released over a range of time scales. Several observational studies have highlighted that the magnetic field evolution comprises three characteristic types of mechanisms that contribute dominantly to the buildup of free magnetic energy in the solar atmosphere. These include: (1) the emergence or cancellation of magnetic flux \citep{zhanghongqi1995,leka1996,chenshibata2000, Sterling2010} (2) shearing motions between opposite polarity magnetic fields \citep{ambastha1993,Demoulin2002_MagHelShMot,Vemareddy2017_SucHom,chintzoglou19}, and (3) the rotation of sunspots (e.g., \citealt{brown2003, tian2006, vemareddy2012_sunspot_rot, Vemareddy2016_sunspot_rot} ). While the evolution of magnetic fields results in energy storage, simultaneous observations reveal distinct pre-eruptive coronal structures such as X-ray sigmoids, H$\alpha$ filaments/prominences, and EUV hot channels. These features are interpreted as twisted magnetic fields formed due to stresses at the foot points of magnetic field lines. The non-potentiality of these twisted fields is quantified using
a variety of measures such as electric currents, magnetic free energy, magnetic helicity, etc. (e.g., \citet{vemareddy2012_sunspot_rot, Pariat2017_HelRat, Zuccarello2018_Threshold_hj}), which, in turn, serves to determine the ability of an AR to produce eruptions. Studying magnetic field evolution makes it essential to understand the mechanisms of twisted flux formation, identify the eruption's stage and trigger, and estimate the amount of stored magnetic energy and other non-potential parameters.

Studying magnetic field evolution requires observations of the magnetic fields in the three-dimensional corona. Since routine coronal magnetic field measurements are not available, extrapolation techniques have been developed to construct 3D magnetic fields based on photospheric magnetic field measurements \citep{Schrijver2008_nlff, WiegelmannSakurai2012}. Typically, these models assume that the coronal magnetic field is force-free, meaning there is a zero Lorentz force. This assumption arises from the low-$\beta$ plasma conditions and the large Alfv\'en speed relative to the photospheric flow speed. Consequently, the evolution of the coronal magnetic field is described as a quasi-static process by a sequence of successive force-free equilibria in response to changes in the surface magnetic field \citep[e.g.,][]{Aly1990_QuasiStat,vemareddy2014_Quasi_Stat}. Additionally, the magneto-friction (MF) method is another technique used to construct force-free coronal magnetic fields, based on a simplified induction equation \citep{Yang1986_MagFric}. Recently, time-dependent, data-driven MF models, with a provision to inject sufficient energy and helicity from the bottom boundary of the computational domain, have been developed to study the long-term evolution of ARs \citep{Cheung2012_MF, Pomoell2019_EleFld, Vemareddy2024_MF_11429} over several days. Compared to static extrapolation techniques, these models have the advantage of capturing dynamic evolution while retaining memory from previous equilibrium stages. 

The objective of this study is to simulate the magnetic evolution of NOAA AR 13500 using a data-driven MF method. This AR was in its decay phase at the time of interest, exhibiting a decreasing magnetic flux content, and produced a fast CME associated with a sigmoidal eruption. In particular, this work aims to capture both the formation and eruption of the sigmoid structure by self-consistently modeling the magnetic field evolution with photospheric vector magnetic field observations. Previous data-driven MF simulations of NOAA AR 11149 have effectively reproduced the formation of twisted flux along the main polarity inversion line (PIL) with an unprecedented morphological agreement with observations \citep{Vemareddy2024_MF_11429}, highlighting the promising capabilities of these models. With such a robust modeling tool, one can assess the variety of non-potential parameters, especially the ratio of the non-potential magnetic helicity and the total relative magnetic helicity \citep{Pariat2017_HelRat, Zuccarello2018_Threshold_hj} for their ability to predict the eruptive potential of an AR. Zero-$\beta$ simulations of \citet{Zuccarello2018_Threshold_hj} suggested that eruptive ARs possess a helicity ratio of approximately 0.29, that is coincident with the onset of torus instability; however, this association requires verification in several  AR cases for a general validity of the helicity threshold. The manuscript is organized as follows. Section~\ref{Obs} presents the observations of the AR and its magnetic evolution leading to the CME. Section~\ref{SimMagEvol} provides the simulation results and their comparison with the observations, along with the characteristic non-potential parameters describing the eruptive potential of the AR. A summary with a brief discussion is given in Section~\ref{SummDisc}.

\begin{figure*}[!ht]
    \centering
    \includegraphics[width=.95\textwidth,clip=]{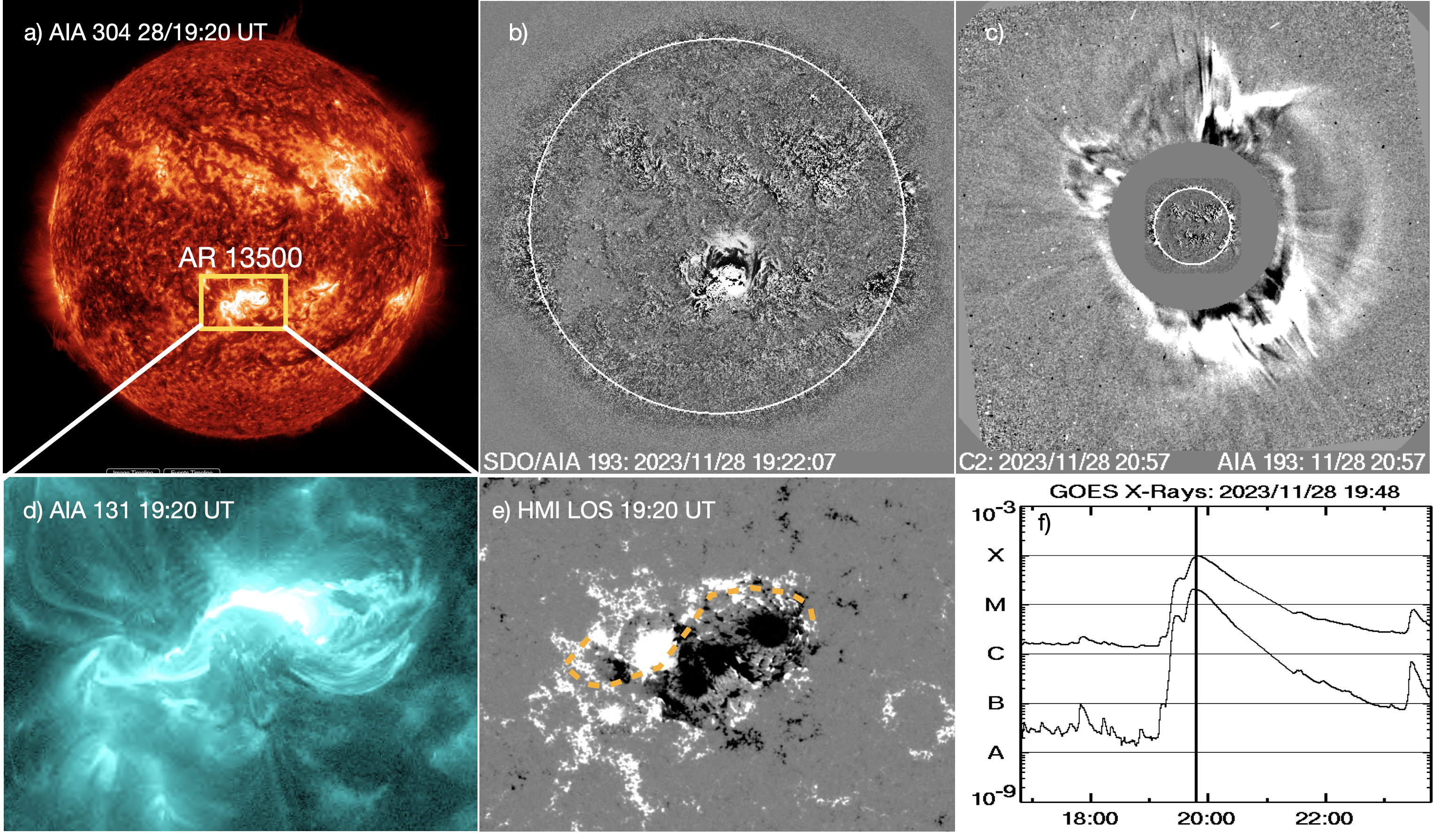}
    \caption{ Multi-observation scenario of the CME eruption from the AR 13500 on November 28, 2023. {\bf a)} Full disk image of the sun in AIA 304~\AA~showing the AR location (yellow rectangular box), {\bf b)} running difference image in AIA 193~\AA, indicating the onset of the eruption from the AR 13500, {\bf c)} LASCO/C2 difference image showing the halo-CME emerging from the sun, {\bf d)} AIA 131~\AA~image of the AR 13500 present with an S-shape sigmoid, {\bf e)} HMI magnetogram displaying the magnetic field distribution in the AR. For comparison, the trace of the sigmoid is overlaid (yellow dashed curve), {\bf f)} GOES X-ray flux light curve referring to the M9.8 flare starting from 28T19:35 UT. }
    \label{fig1}
\end{figure*}

\section{Observations and analysis}
\label{Obs}
We studied the magnetic origins of the CME that occurred on November 28, 2023, at 19:50 UT. This CME was identified as the solar source of an intense geomagnetic storm on December 1, which began at 01:00 UT and peaked at -105 nT around 12:00 UT the same day. 
\subsection{The CME and its source region}
Observations of the source region were obtained using instruments aboard the Solar Dynamics Observatory (SDO). The Atmospheric Imaging Assembly (AIA; \citealt{Lemen2012}) continuously monitors the full solar disk in seven EUV wavelengths, capturing dynamical processes from the chromosphere to the corona with a 12-second cadence and 0.6 arc-second per pixel resolution. Magnetic field measurements at the photosphere were obtained using the Helioseismic and Magnetic Imager (HMI; \citealt{schou2012}), which has a pixel resolution of 0.5 arc-second and a cadence of 12 minutes. White light observations of the extended corona up to 32 R$_\odot$ are obtained from the LASCO instrument onboard SoHO. Figure~\ref{fig1}(a) displays the full-disk image of the Sun in the AIA 304~\AA~waveband, showing AR 13500 located near the disk center ($S20^{o}W04^{o}$) on November 28. The coronal emission is structured, featuring an S-shaped sigmoidal morphology (Figure~\ref{fig1}(a,d,e)). The onset of the eruption (CME launch) occurred at around 19:15 UT as seen by AIA 193 \AA~running difference images, displayed in Figure~\ref{fig1}(b). The CME started emerging in LASCO/C2 FOV from 20:12 UT as a full-halo event, which can be noticed in Figure~\ref{fig1}(c) of the running difference LASCO/C2 image. This eruption was associated with an M9.8 solar flare that began at 19:35 UT and peaked at 19:50 UT (Figure~\ref{fig1}(f)). The CME's linear speed, as measured in the LASCO field of view, was 741 km/s. The travel time of this CME to Earth was $\approx$53 hours. In the subsequent sections, we analyze the magnetic evolution of the AR that led to the formation and eruption of the sigmoidal structure and employ data-driven simulations to gain insights into the mechanisms behind the eruption.

\begin{figure*}[!ht]
\centering
\includegraphics[width=0.95\textwidth, clip=]{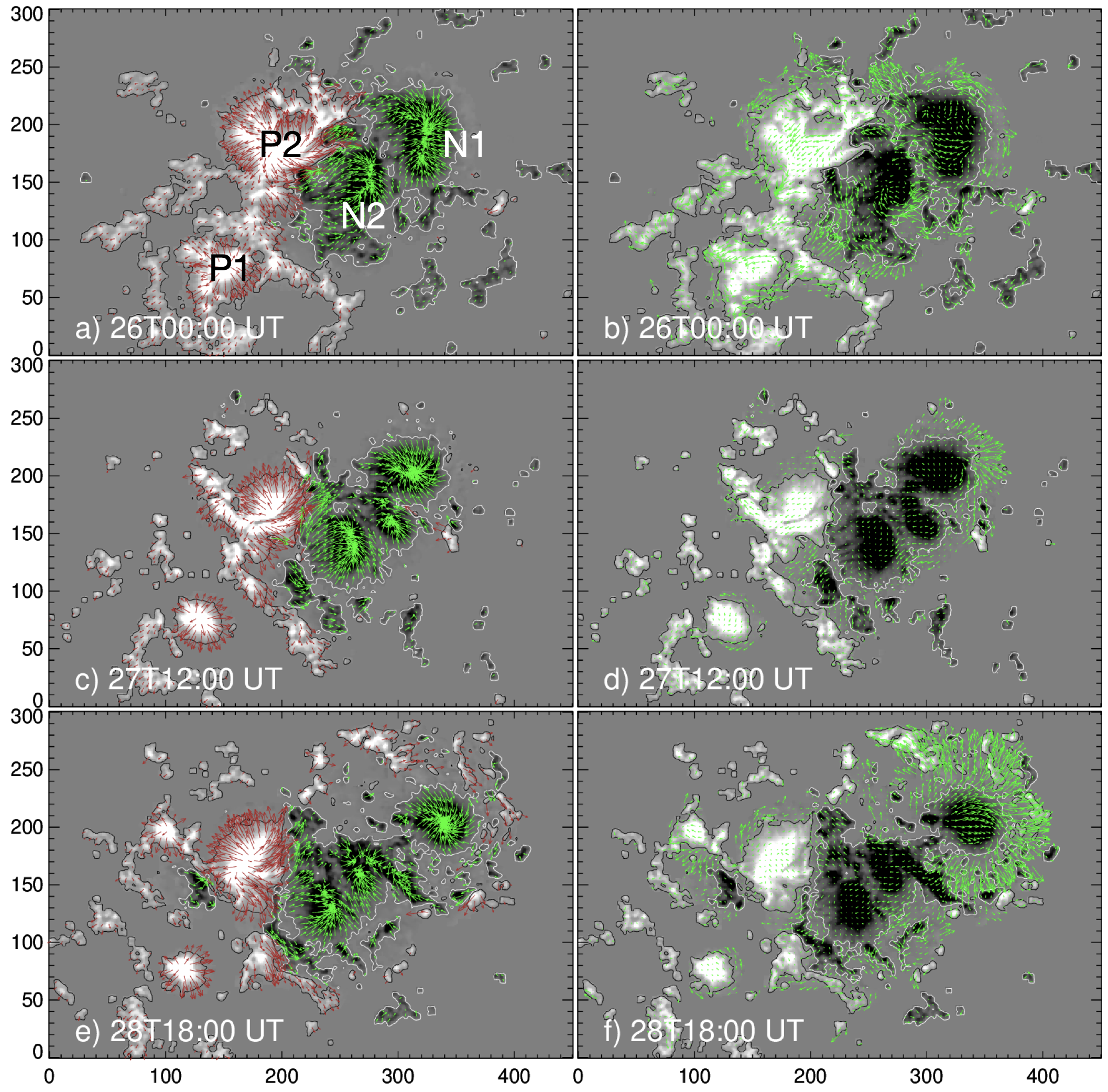}
\caption{{\bf Left:} HMI vector magnetograms of the AR 13500 at different epochs of its evolution. {\bf Right:} Derived velocity of the flux motions. Arrows referring to horizontal field are over plotted on the vertical component of the magnetic field. Prominent polarities are labeled in panel a).  Notice that while N2 and P2 move antiparallel, N1 separates from N2 with a proper motion. Axis units are in pixels of 0.5 arcsec size.}
\label{fig_vmg}
\end{figure*}

\subsection{Magnetic evolution of the AR}
Figure~\ref{fig_vmg} displays observations of the vector magnetic field in the photosphere of the AR during different stages of its evolution. The overall polarity distribution of the AR follows Hale's polarity law for solar cycle 24, with the leading polarity being negative and the following polarity positive in the southern hemisphere. Notably, the AR contains inner opposite polarity regions, labeled N2 and P2, as well as outer polarity regions of the same orientation, labeled N1 and P1, as observed on November 26. The inner pair creates a polarity inversion line, which generates a sheared magnetic field as the regions move slowly in opposite directions over time. Meanwhile, the N1 region, initially close to the N2 region on November 26, experiences proper motion and becomes detached by the end of November 28. 

\begin{figure*}[!ht]
    \centering
    \includegraphics[width=.9\textwidth,clip=]{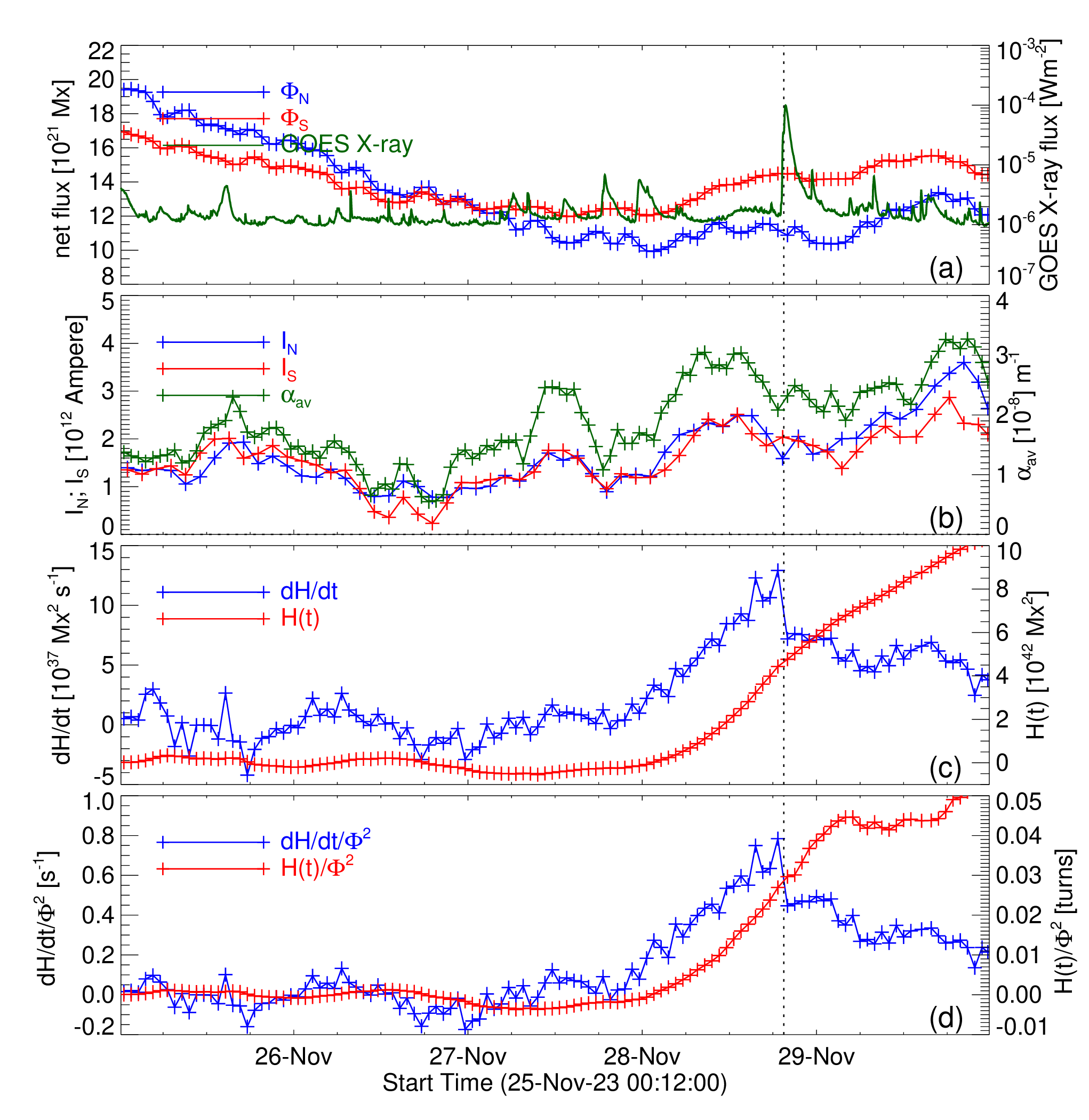}
    \caption{ Time evolution of non-potential parameters in the AR 13500, {\bf a)} Net magnetic flux showing a decreasing flux content as it evolves in time interacting magnetic polarities. GOES flux is also shown with y-axis scale on the right., {\bf b)} Net vertical current in each polarity flux region, along with average twist parameter ($\alpha_{av}$), {\bf c)} the helicity injection rate ($dH/dt$, Y-axis on left) and accumulated helicity $H(t)$ (red, Y-axis on right). As marked by vertical dotted line, the peak $dH/dt$ is co-temporal with the CME eruption and the M9.8 flare {\bf d)} helicity flux normalized with the square of magnetic flux. }
    \label{fig_mag_evol}
\end{figure*}

Importantly, the sizes of the polarities decrease over the two days of evolution, which is reflected in the decreasing net magnetic flux shown in Figure~\ref{fig_mag_evol}(a). The underlying flux motions are key to the injection of helicity and magnetic energy into the AR \citep{vemareddy2012_hinj, Vemareddy2015_HelEne}. Due to these slow footpoint motions, the magnetic fields are stressed and develop largely field-aligned electric currents, which can be seen in the evolution of the net electric current shown in Figure~\ref{fig_mag_evol}(b) in each polarity ($I_\mathrm{N}$ and $I_\mathrm{S}$, respectively) towards the eruption time. As a result, the net average force-free twist parameter $\alpha_{av}$ shown in the same panel) also exhibits an increasing trend. 

Using a time sequence of the vector magnetic fields, we first derive the vector velocity, $\mathbf{V}$, of the flux motions using the DAVE4VM \citep{schuck2008} technique, and then  compute helicity flux injection from the photospheric surface \citep{Berger1984} as
\begin{equation}
    \left. \frac{dH}{dt} \right|_S = 
2 \int_S (\mathbf{A}_p \cdot \mathbf{B}_t) V_{n} \, dS 
- 2 \int_S (\mathbf{A}_p \cdot \mathbf{V}_{t}) B_n \, dS.
\label{eq_hinj}
\end{equation}
As shown in Figure~\ref{fig_mag_evol}\cm{(c)}, $dH/dt$ increases predominantly with a positive sign from early November 27 to the time of eruption which is linked to the polarities in shearing motion. The $dH/dt$ reaches around $3\times 10^{37}$ Mx$^2$/s towards the end of November 27, following a phase of increasing positive-sign helicity injection.  Especially, the peak $dH/dt=13\times10^{37}$ Mx$^2$/s occurs at 18:00 UT on November 28 (vertical dotted line), which is co-temporal with the onset time of the CME and the associated M9.8 flare. The leading N1 polarity exhibits separating motion from November 27, the jump in $dH/dt$ at the time of eruption likely corresponds to detached N1 from N2.  By the time of eruption, the AR has accumulated a net helicity of around $5 \times 10^{42} Mx^2$, which is significant for an eruptive potential.

\begin{figure*}
    \centering
    \includegraphics[width=.7\textwidth,clip=]{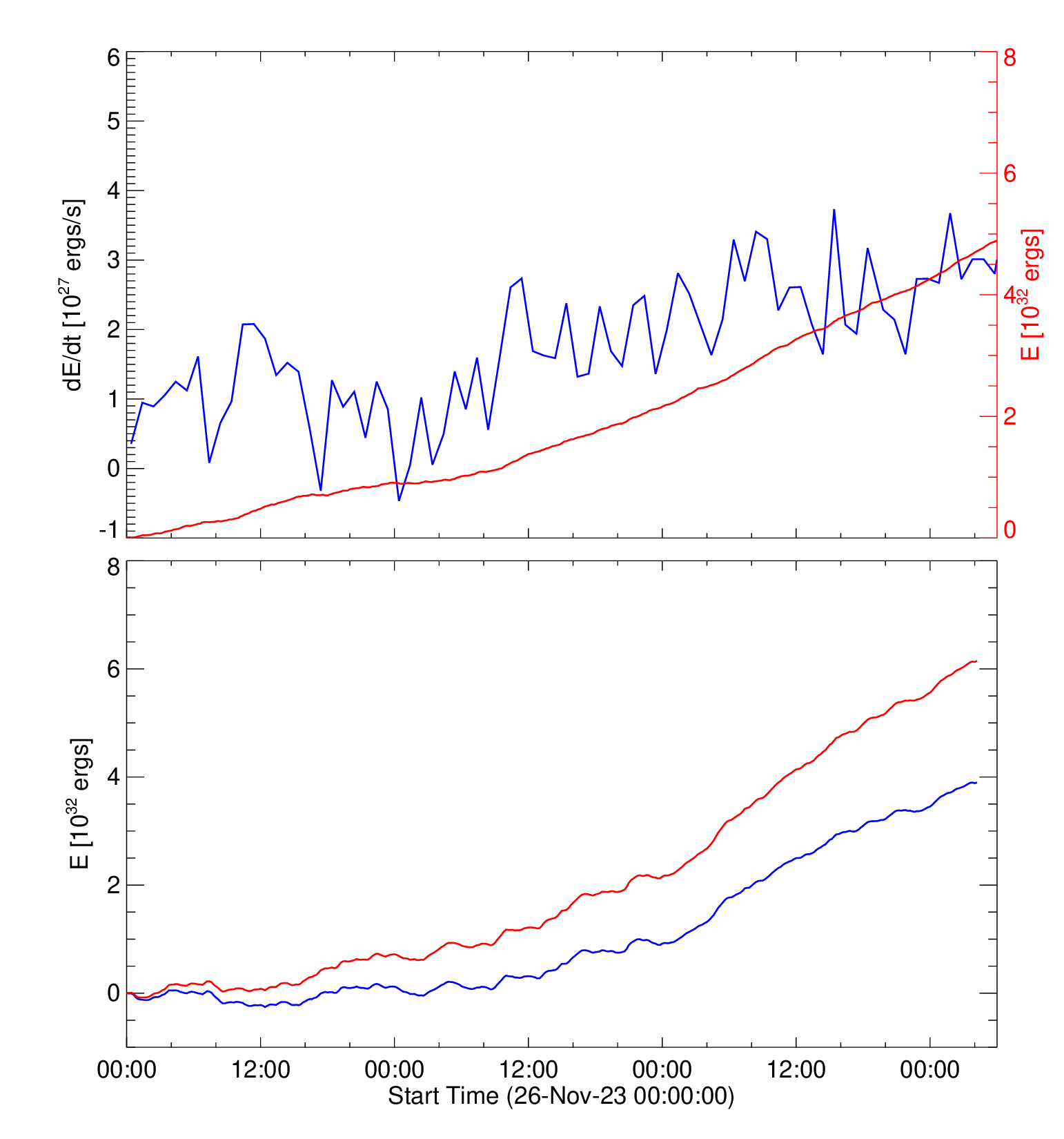}
    \caption{Constraining the energy injection for MF simulations from vector field observations. {\bf Top:} Computed Poynting flux ($dE/dt$) and its accumulated quantity from the observed magnetic field at the photospheric surface. {\bf Bottom:} Accumulated poynting flux from the electric fields derived with $U=120$ m/s (blue) and $U=150$ m/s (red).}
    \label{fig_poyflx}
\end{figure*}
Since helicity is proportional to the square of net magnetic flux \citep{Berger1984},  we express an average amount of AR twist injection by normalising both $dH/dt$ and $H$ by the square of the magnetic flux, $\Phi^2$ Figure~\ref{fig_mag_evol}(d). By the time of the CME onset, $H/\Phi^2$ has reached a value of 0.04 turns. This quantity is limited up to 0.02 turns in ARs that produce confined flares, but has been found to be as high as 0.15 turns during six days of evolution in successively erupting ARs reported in 
\citet{Vemareddy2017_SucHom, vemareddy2019_VeryFast}.  

\section{Simulating the magnetic evolution of the AR}
\label{SimMagEvol}
We employ a data-driven magnetofriction (MF) model to simulate the magnetic evolution of the AR, as detailed in the
recent works \citep{Cheung2012_MF,Pomoell2019_EleFld, Vemareddy2024_MF_11429}. We implement this simulation model in the PENCIL code, with an additional special driver module to supply observed magnetic fields at the bottom of the computational grid while the simulation advances in time \citep{Vemareddy2024_MF_bb}. The initial field is a potential field, which is then driven by electric fields that are derived from time-varying photospheric vector magnetic field observations \citep{Fisher2010_EstEleFld}. In order to inject sufficient helicity and energy into the coronal field, the electric field is additionally supplied with a non-inductive contribution controlled by an ad-hoc parameter, $U$. 

\begin{figure*}[!ht]
    \centering
    \includegraphics[width=.95\textwidth,clip=]{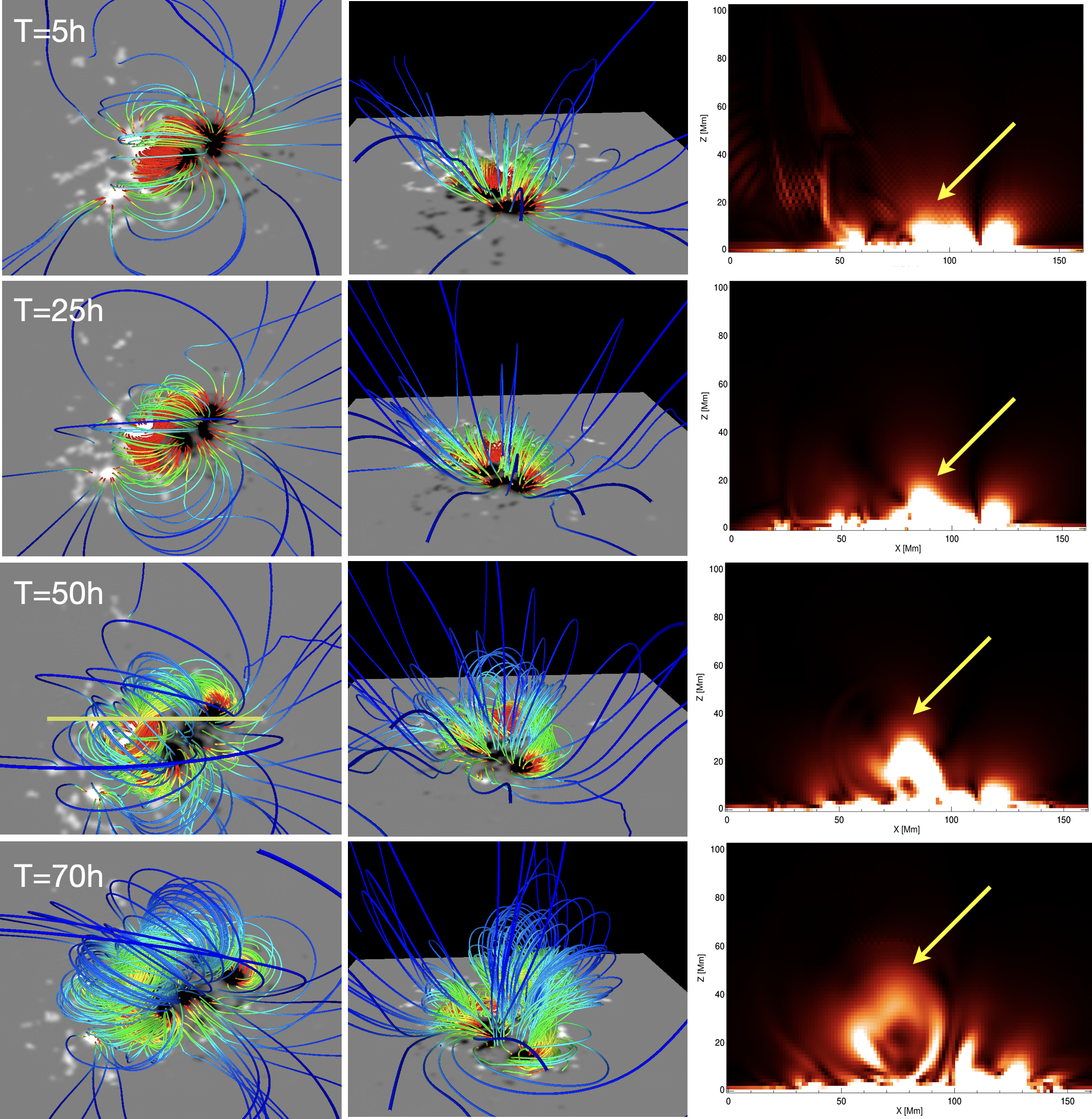}
    \caption{ The evolution of AR magnetic structure at different epochs of the simulation. {\bf first column:} top view, {\bf second column:} perspective view, {\bf third column:} Total current in the vertical cross section placed across the sigmoid.  Magnetic field lines are colored by their field strengths, and the background image is the normal magnetic field Bz at the bottom of the computational domain ($z = 0$). The simulation captures the formation of the twisted flux rope by 50th hour along the PIL and its slow rise motion later. Arrow points to twisted flux that is being formed and rises progressively.}
    \label{fig_sim_mos}
\end{figure*}

Figure~\ref{fig_poyflx} plots the Poynting flux, $dE/dt$, derived from the direct magnetic field and velocity field (top panel) in comparison with that derived from the electric field being used as the driver field (bottom panel). Starting from November 26, the total magnetic energy, $E$, accumulated by the time of the eruption (28T19:12 UT; 67 hours of evolution) is $4\times10^{32}$ ergs, which is to be constrained in the simulation through the driver electric field. As shown in the plot, the electric field derived with an ad-hoc parameter of $U=150$ m/s better represents the energy injection scenario in the observations, which is being used to simulate the coronal field evolution. 

\begin{figure*}[!ht]
\centering
\includegraphics[width=0.95\textwidth,clip=]{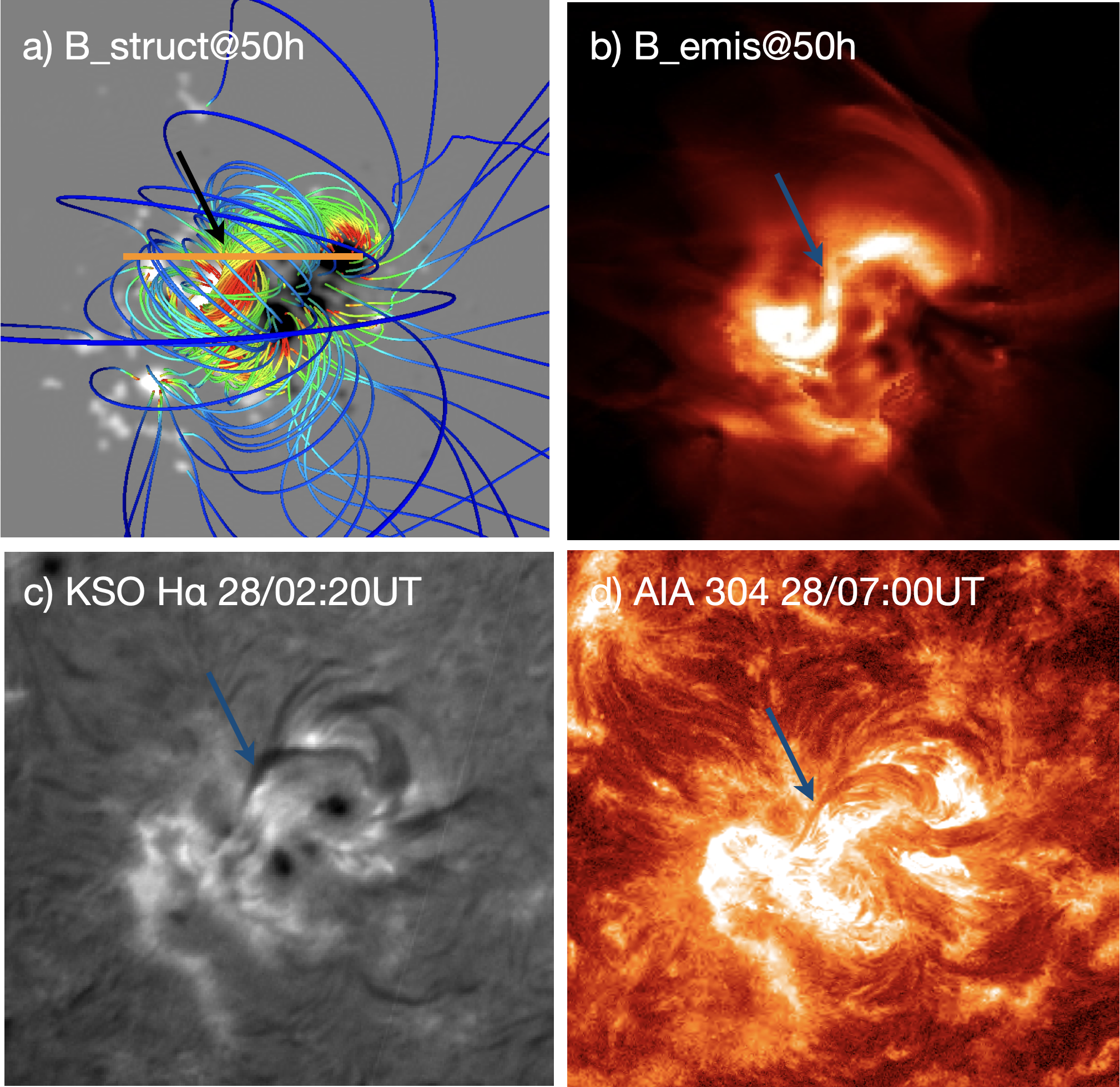}
    \caption{Comparison of simulated magnetic structure with the coronal observations; {\bf a)} rendered magnetic structure at 45h of simulated field, {\bf b)} proxy emission map of the simulated magnetic field, {\bf c)} AIA 304~\AA~ image at 28T07:00 UT. Note that the striking morphological similarity of the simulated proxy emission with the AIA 304~\AA~image broadly reproducing the sigmoidal feature. {\bf d)} H-$\alpha$~ image from KSO at 28T02:20 UT. Blue arrow points to the filament and its supporting magnetic structure. }
\label{fig_mag_comp}
\end{figure*}

In the first two columns of Figure~\ref{fig_sim_mos}, snapshots of the simulated magnetic field are displayed at different times. To visualize the twisted fields, we trace field lines at locations of strong electric currents in the lower part of the computational domain. The simulation begins with the potential field (PF) constructed from the radial component of vector magnetic field observations at 26T00:00 UT. The initial PF comprises an inner arcade along the PIL, which is enveloped by an outer arcade that connects the peripheral parts of opposite polarities. This configuration evolves slowly with the changing boundary conditions, still largely resembling a PF until about 25 hours into the simulation. This behavior is consistent with the energy being injected from the bottom boundary (see Figure~\ref{fig_poyflx}). As energy and helicity accumulate, further evolution transforms the inner arcade into a sheared arcade, which eventually develops into a highly twisted structure (i.e., into an FR) resembling a sigmoid, as observed in the snapshot at the 50th hour of the simulation, corresponding to the end of November 27. In the third column of Figure~\ref{fig_sim_mos}, we present the total current ($|\mathbf{J}|$) distribution in a vertical cross-section that is placed across this twisted structure (FR). These panels illustrate the formation and slow ascent of the FR over the three days of the simulation. We observe that the top of the FR reaches a height of around 50 Mm by the 50th hour and extends to almost 80 Mm by the 68th hour  (the observed eruption time),  which is a part of quasi-static evolution. With the driving boundary conditions, the FR builds by developing strong electric currents, and the Lorentz force overcomes the tension of the overlying field. As a result, the FR experiences an upward slow rise motion as seen in this quasi-static phase of evolution. Observations indicate that pre-eruptive structures, such as prominences and EUV hot channels, are found at a similar heights in the corona (e.g., \citealt{zhangj2012, Vemareddy2022_hotchan}). 

\begin{figure*}
    \centering
    \includegraphics[width=0.95\textwidth,clip=]{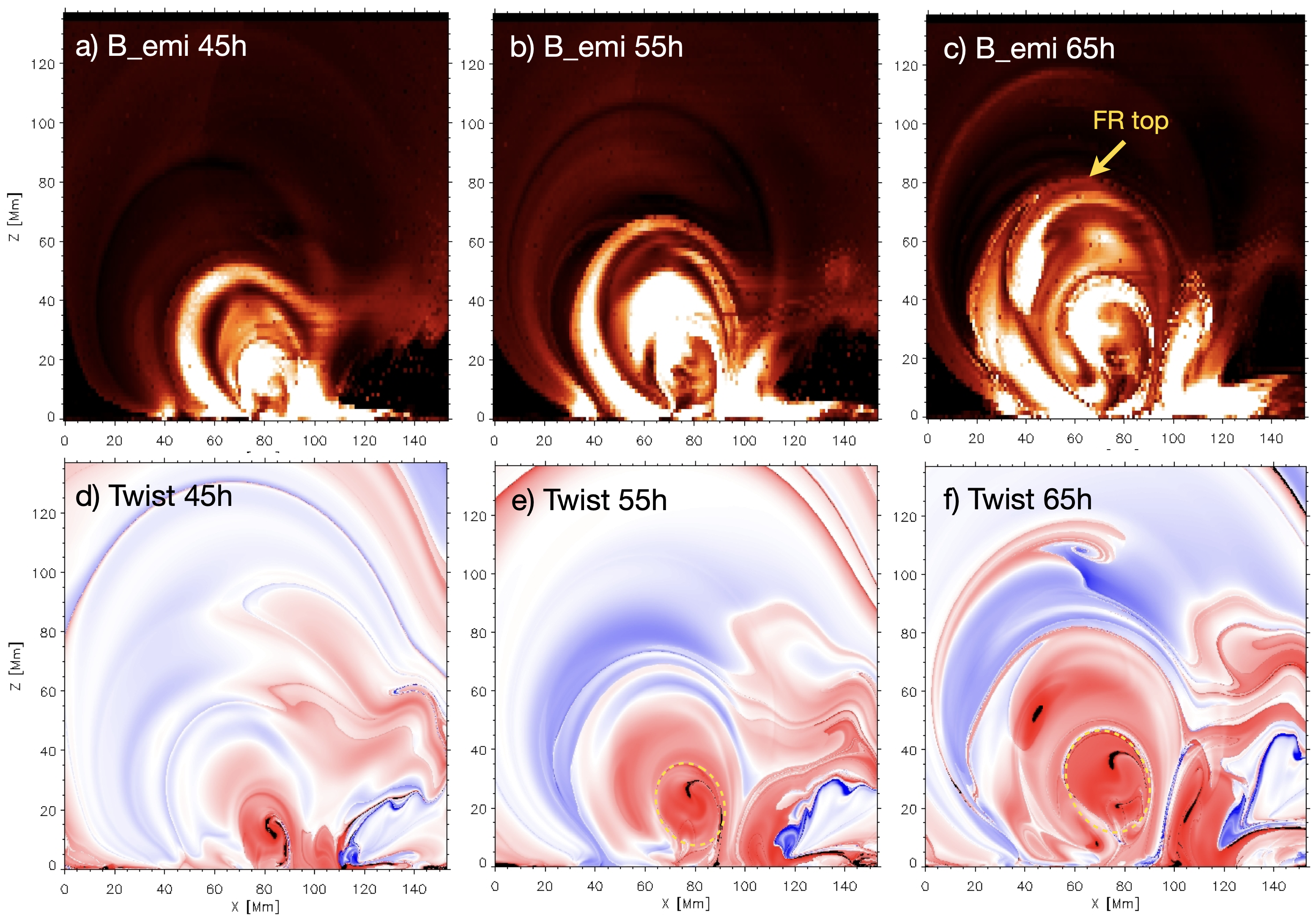}
    \caption{Buildup and onset of flux rope eruption. {\bf a-c)} proxy emission in the vertical slice (orange line in panel Figure~\ref{fig_mag_comp}a ) at different epochs, showing the slow upward moving FR with its top reaching 80 Mm by 65h. {\bf d-f)} twist maps in the vertical slice. Red (blue) color indicates positive or right handed twist scaled to 2 turns. Core of the FR is indicated with yellow oval within which the average field line twist increases from 1.28 turns at 55h to 1.43 turns at 65h.   
}
\label{fig_tst}
\end{figure*}

In Figure~\ref{fig_mag_comp}, we compare the simulated magnetic structure with the coronal observations at the 45th hour, when the FR has formed. We note, at this point, that the transition from a sheared arcade to an (initially weakly twisted) FR is a gradual process, so it cannot be associated with a single point in time \citep[e.g.,][]{patsourakos20}. However, visual inspection of field lines and the twist maps described below suggests that a well-developed FR is already present at this time. The comparison is done by generating a proxy emission map, which is  obtained from integrating along the line of sight (here, in a top-down view) the average value of the square of the total current density, $J^2$, over magnetic field lines (see details in \citet{Cheung2012_MF}). As can be seen, the FR is co-spatial with the sigmoidal feature in the proxy emission map. This sigmoidal feature has a remarkable morphological similarity with the one observed in the AIA 304\AA~image, and H$\alpha$ image from the Kodaikanal Solar Observatory (KSO), especially regarding the orientation and curvature of its lobes. This indicates that our simulation successfully captures 
important characteristics of the coronal magnetic field in the AR. 

\begin{figure*}
    \centering
    \includegraphics[width=0.8\textwidth, clip=]{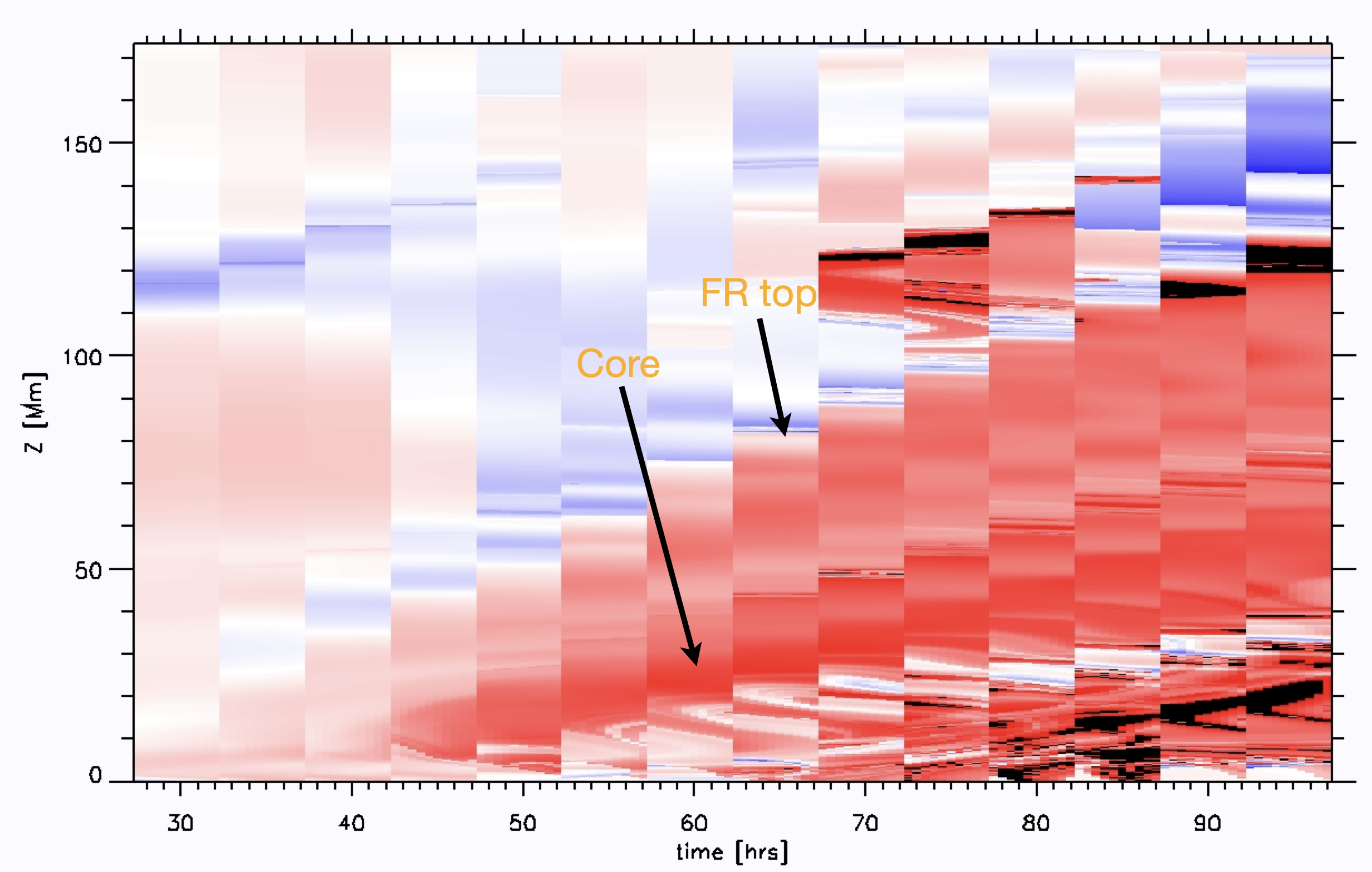}
    \caption{Space-time map of twist along the flux rope height, in a plane as shown in Figure~\ref{fig_tst}. Red color refers to positive twist in the core, which is rising in time from 50 hours onwards. Note that until 45 hours, the flux rope is forming, so weaker twist (less intense red) is realized. }
    \label{fig_ht_tst}
\end{figure*}

To illustrate the slow ascent of the FR mentioned above, the proxy emission is extracted along a slice plane that captures the cross section of the FR, as depicted in Figure~\ref{fig_tst}(a–c). The field line twist ($T_w$) is also computed in the same plane with the expression given by \citep{Berger2006, LiuRui2016}
\begin{eqnarray}
{T}_{w}=\int_{L}\displaystyle \frac{{\mu }_{0}{{\mathbf{J}}}_{||}}{4\pi B}\,dl=\int_{L}\displaystyle \frac{{\rm{\nabla }}\times {\mathbf{B}}\cdot {\mathbf{B}}}{4\pi {B}^{2}}\,dl.
\end{eqnarray}
as shown in Figure~\ref{fig_tst}(d–f). These panels clearly illustrate that the FR rises gradually, with its top reaching a height of approximately 80 Mm by the 65th hour. It is noteworthy that the maximum twist within the FR core (enclosed by the yellow oval) is about 3.2 turns, while the average twist increases from 1.28 turns at 55 hours to 1.43 turns at 65th hour, indicating a development of the helical kink instability according to \citep{torok2004}, that could have initiated the onset of the eruption. Note that the core of the FR is around 40 Mm.

To better assess the evolution of the slow FR rise, we prepared a stack of vertical slits from each plane cut of the FR. For clarity, from Figure~\ref{fig_tst} a slit is extracted at around 80 Mm along the $X$-direction, from which a twist map was produced and presented in Figure~\ref{fig_ht_tst}. Red color is a positive twist of the FR, the more intense red corresponds to the core. The FR's gradual ascent becomes evident after its formation at around 45 hours, revealing a slow-rise motion as part of its further build up and quasi-static evolution. By 65th hr, the FR (top, point by arrow) reached a height of 80 Mm, and it rises further until the end of simulation at 96 hrs. Note that the driver is switched off at 70 hrs, just after the observed eruption at the 67th hour. The rise motion thereafter, while slow, mimics the FR eruption. It is important to mention that a transition from slow rise motion until 67 hours to the later time is not present as we usually observe during the initiation of the eruption \citep{Vasantharaju2018}, which is a drawback of MF simulations as the velocity is controlled by the constant frictional parameters.

\begin{figure*}[!ht]
    \centering
    \includegraphics[width=.7\textwidth,clip=]{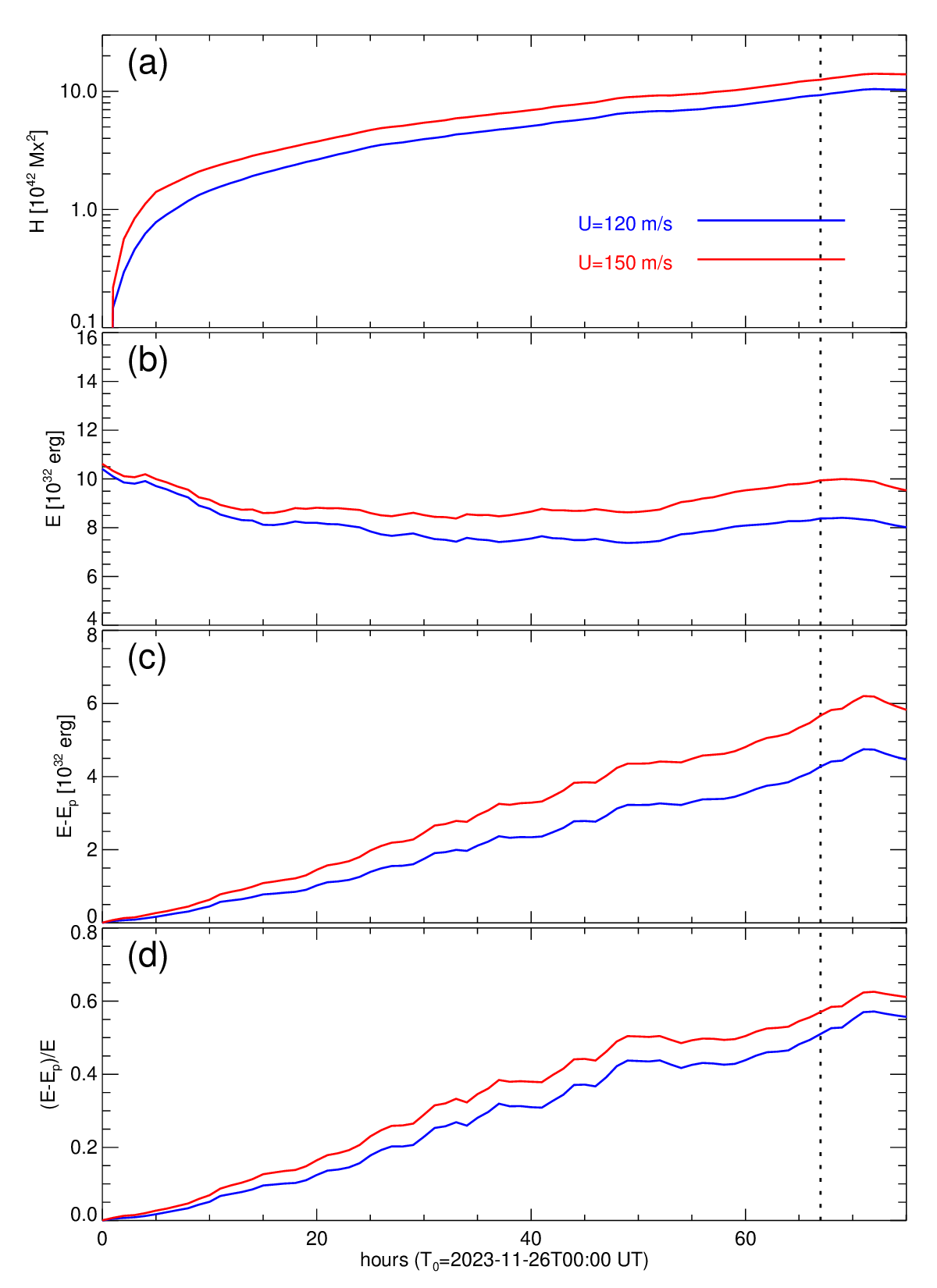}
    \caption{Time evolution of energy and helicity parameters in the computation volume. (a) Total magnetic helicity, (b) total magnetic energy, (c) free magnetic energy, and (d) fractional free energy with respect to the total energy. Red, and blue correspond to the runs with $U=120$ m/s, and $U=150$ m/s, respectively. The vertical dotted line refers to the eruption time at 19:00 UT on March 28, at which time the free-energy fraction is 0.51, 0.57.}
    \label{fig_tevol} 
\end{figure*}

Figure~\ref{fig_tevol} presents the time evolution of the computed relative magnetic helicity (H) and the magnetic energy (E) of the simulated magnetic field. The volume helicity (H) steadily accumulates from the start of the simulation, reaching a value of \(10\times10^{42} \,\text{Mx}^2\) for the run using an electric field derived with \(U=120\,\text{m/s}\). This value increases slightly to \(14 \times 10^{42} \, \text{Mx}^2\) for the run with \(U=150\,\text{m/s}\). Meanwhile, the total magnetic energy \(E\) decreases slightly from \(10.5\) to \(8 \times 10^{32}\,\text{ergs}\) over three days of evolution. However, the free magnetic energy \(E_f=E-E_p\) continuously increases, reaching \(4\times10^{32}\,\text{ergs}\) in the case of the \(U= 120\,\text{m/s}\) run. This observation can be explained by the fact that the AR is in a decay phase, resulting in a net decrease in magnetic energy during this time period. This example highlights that total magnetic energy \(E\) is not a reliable indicator of the eruptive potential of an AR. Instead, the free magnetic energy \(E_f\) which is associated with the non-potential field, plays a crucial role and varies under different circumstances. At the time of the observed eruption, the fractional free energy ($E_f/E$) is 51\% for the \(U=120\,\text{m/s}\) run and 57\% for the \(U=150\,\text{m/s}\) run. According to previous studies \citep{Pomoell2019_EleFld,Vemareddy2024_MF_11429} the latter run appears to have a higher potential for the eruption.

\subsection{The $H_J/H_V$ ratio}
\begin{figure*}[!ht]
\centering
\includegraphics[width=.7\textwidth,clip=]{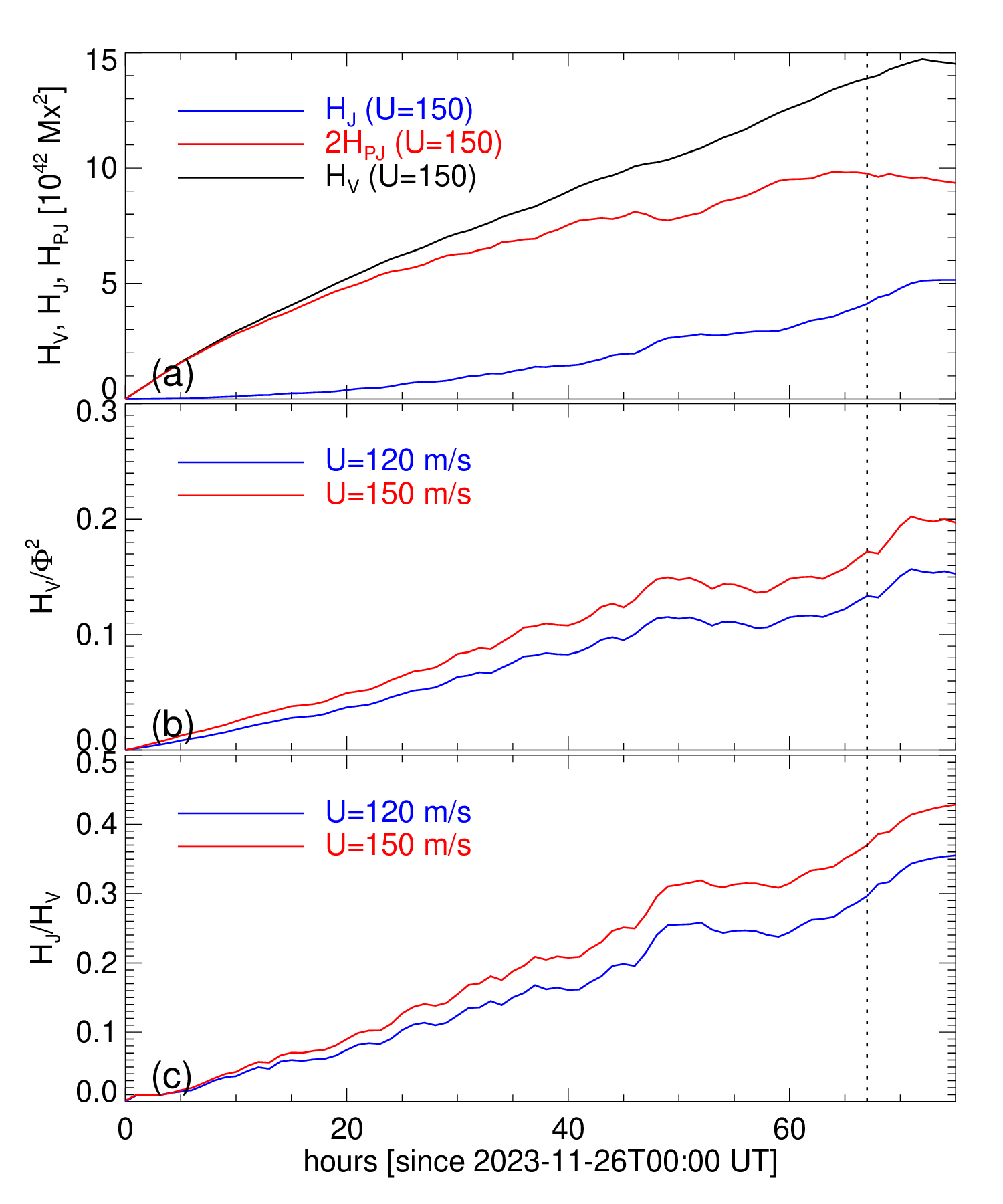}
\caption{a) Time evolution of $H_V$, $H_J$, $H_{PJ}$ for the $U=150m/s$, b) $H_V/\Phi^2$, c) the ratio $H_J/H_{V}$ for runs with $U=120$, 150 m/s. Vertical dotted line corresponds to CME onset time, when the ratio $H_J/H_V$ is 0.3 and 0.37 for the two runs.}
\label{fig_hjhv} 
\end{figure*}

According to \citet{Berger2003_Topquant}, the relative magnetic helicity can be decomposed as $H_V=H_J + 2 H_{PJ}$, with
\begin{eqnarray*}
H_J&=&\int_V (\mathbf{A}-\mathbf{A}_P)\cdot (\mathbf{B}-\mathbf{B}_p) dV \\
H_{PJ}&=&\int_V \mathbf{A}_p \cdot  ( \mathbf{B}-\mathbf{B}_p ) dV 
\end{eqnarray*}
Here $H_J$ is the magnetic helicity of the non-potential or current-carrying component of the magnetic field $\mathbf{B}_J=\mathbf{B}-\mathbf{B}_p$ and $H_{\rm pJ}$ is the mutual helicity between $\mathbf{B}_p$ and $\mathbf{B}_J$. We evaluate these quantities in the entire simulation box as shown in Figure~\ref{fig_hjhv}(a). As the simulation progresses, both $H_{\rm PJ}$ and $H_J$ accumulate steadily with pumping energy from the bottom boundary. At the time of the observed eruption, they reached $9.76\times10^{42}Mx^2$ and $4.12\times10^{42}Mx^2$, respectively. Since the bottom boundary's $H_V$ is scaled by the square of the net average flux ($\Phi$), we calculated $H_V/\Phi^2$, as shown in Figure~\ref{fig_hjhv}(b). Based on this, the average magnetic field twist at the recorded eruption time is 0.13 and 0.17 turns for $U=120$ and 150 m/s runs, respectively, which are comparable with the reported values of \citet{Zuccarello2018_Threshold_hj}.

We also analyze the evolution of the ratio $H_J/H_V$ which has been shown to have a characteristic pattern that distinguishes eruptive from non-eruptive ARs \citep{Pariat2017_HelRat,Zuccarello2018_Threshold_hj}. This ratio exhibits an increasing trend over time, as illustrated in Figure~\ref{fig_hjhv}(c). At the time of the eruption, $H_J/H_V$ is observed to be 0.3 and 0.37 respectively, for the two runs. Previous studies have investigated time series of nonlinear force-free extrapolations, indicating that the evolution of $H_J/H_V$ differs between eruptive and non-eruptive ARs where an AR is found to erupt when $H_J/H_V>0.1$ \citep{Gupta2021_HelEne_ConEru_flares}.

\begin{figure*}[!ht]
\centering
\includegraphics[width=.7\textwidth,clip=]{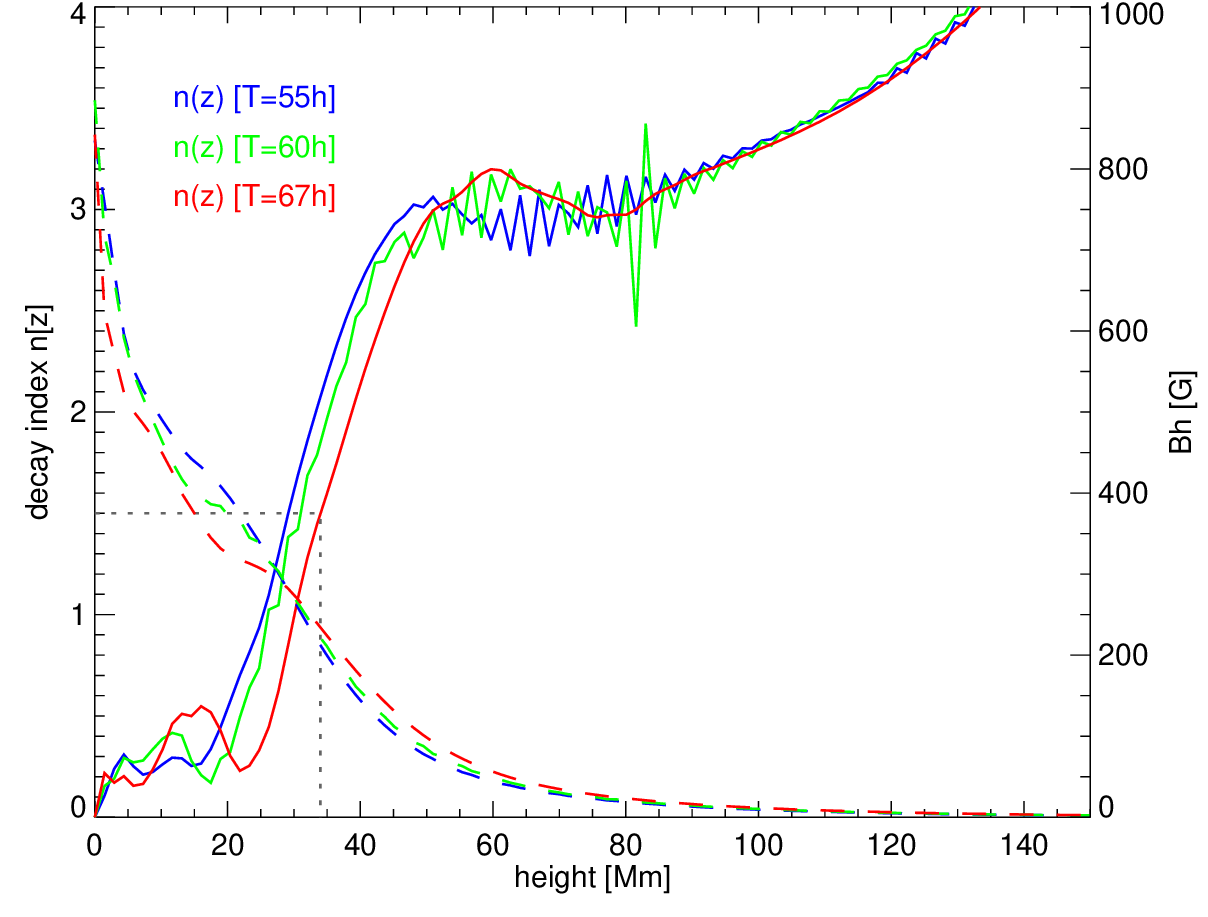}
\caption{Decay index of the simulated magnetic field at {\bf 55th (blue), 60th (green), 67th (red)} hours. Horizontal magnetic field is also plotted with Y-axis on right. Vertical dotted line corresponds to n=1.5 at 34$\pm$1 Mm. }
\label{fig_di} 
\end{figure*}

\subsection{Evidence for torus instability}

Additionally, using a series of line-tied MHD simulations, \citet{Zuccarello2018_Threshold_hj} showed that at the onset of the torus instability the ratio $H_J/H_V\approx0.29$ proposing it as a reliable proxy for identifying the instability threshold. This result, however, comes with an important caveat: the relative helicity $H_V$ is not an additive quantity, and values of the helicity ratio computed over different volumes or with different boundary conditions can vary. In our study, both $H_J$ and $H_V$ are evaluated over the entire simulation domain, allowing the ratio to be interpreted meaningfully in the context of eruptive behavior in the AR. Since the simulated field is quasi-static, without fast eruptive signatures, we considered the observed time of eruption (67th hour) to assess the eruptive behavior of the magnetic structure.

Our results are consistent with these simulations, as we find 
$H_J/H_V\geq0.3$ at the time of eruption. Comparable high values were also reported by \citet{Green2022_HelEvo11158}, who analyzed AR 11158 using non-linear force-free magnetic field extrapolations. While a condition of $H_J/H_V\geq0.1$ has been suggested as sufficient for an AR to be susceptible to eruption, the onset of the torus instability appears to require larger values. This indicates that a threshold of 
$H_J/H_V\geq0.3$ may be applicable for certain ARs, including the case examined here.
In our TMF simulation, the ratio reaches 0.1 once the twisted FR is well developed, after approximately 50 hours of evolution. Thereafter, it increases steadily, and by the time of the observed eruption at around the 65th hour, the FR apex reaches a height of about 80 Mm (with the FR core center located near 32 Mm; see Figure~\ref{fig_tst}), signaling a strong eruptive potential. The subsequent rapid eruption phase is not captured in our simulations, as MF models are inherently limited by the imposed frictional coefficient.

To support this eruption scenario with the determined helicity ratio, we assess the torus criteria $n=-\frac{z}{B_h}\frac{\partial B_h}{\partial z}$ where $B_h=\sqrt{B_x^2+B_y^2}$ is the horizontal field strength as a function of height z. Figure~\ref{fig_di} presents the decay index $n$ of the simulated magnetic field at the 55th (blue), 60th (green), and 67th (red) hours. At these times, $n$ exceeds 1.5 at heights of 29 Mm, 30.5 Mm, and 33.5 Mm, respectively. This behavior indicates a rapidly decreasing overlying magnetic field, while the critical height increases as the flux rope (FR) gradually rises. However, at 67 hours-evident from Figure~\ref{fig_tst}-the FR core (outlined by the yellow oval) has its center located above 30 Mm and its top near 45 Mm. Therefore, the observed eruption time (67 hours) is more appropriate for associating the torus instability with the corresponding $H_J/H_V$ ratio. This observational result further suggests that, once the torus instability criterion is satisfied, the helicity ratio (0.3) at the time of the observed eruption may represent an upper limit for triggering the FR eruption. Moreover, in the case of capturing transition from slow quasi-static to fast rise motion, the eruption time is more accurately determined for a reliable basis for comparing $H_J/H_V$ with the critical decay index $n_{crit}$. By self-consistently modeling the AR magnetic field evolution, this study strengthens the evidence for an association between the torus instability and the threshold ratio, while highlighting the need for further case studies to establish its general validity.

\section{Summary and Discussion}
\label{SummDisc}
We investigated the magnetic origin of the CME that occurred on November 28, 2023, at 19:50 UT from AR 13500 located near the solar disk center. This event was accompanied by an M9.8 flare and launched a CME toward Earth at a linear speed of 741 km/s, whose encounter with Earth triggers an intense geomagnetic storm (-105 nT) on December 1, 2023. The CME was associated with an S-shaped sigmoidal structure supported by inner polarity regions (N2, P2) forming the main sheared PIL through persistent shear motions, while the outer polarities (N1, P1) evolved via proper motion (See Figure~\ref{fig_vmg}). Over the three days of the AR magnetic evolution (November 26–28), the polarities shrank, leading to a decline in net magnetic flux, while flux motions injected significant magnetic helicity and energy into the coronal field, accompanied by increasing electric currents and an increasing force-free twist parameter (\(\alpha_{av}\)). Notably, the helicity flux peaked at 18:00 UT on November 28, coinciding with the CME onset at 28T19:50 UT, implying the key role of helicity injection in triggering eruptions. By this time, the AR had accumulated a net helicity of \(5\times10^{42}\) Mx\(^2\) and a normalized helicity twist (\(H/\Phi^2\)) of 0.04 turns, which is above the threshold for confined-flare ARs (\(\leq0.02\) turns) but below the levels in highly eruptive cases (\(\approx0.15\) turns) suggesting a moderate-to-strong eruptive potential of the AR \citep{Vemareddy2017_SucHom}.

By employing data-driven MF model, we simulate the magnetic evolution of the AR since 2.8 days in advance of the CME eruption. The energy injection for the simulation is controlled by an ad hoc parameter that is at $U=150$m/s best matches the observed energy injection, yielding a total magnetic energy of $4\times10^{32}$ ergs by eruption time. From the initial potential field configuration, the AR evolves gradually into a sheared arcade and further as a twisted flux rope, mimicking a sigmoid configuration. Proxy emission maps exhibit a strong morphological agreement between simulated and observed sigmoidal features, which is an indication of the right amount of twist in the coronal magnetic field. The FR structure rises slowly from the point of its formation and reaches 80 Mm height by the time of its eventual eruption ($T=67$ hours). In addition, the average field line twist in the core of the twisted FR reached 1.43 turns, suggesting its critical state for ideal kink instability \citep{torok2004}. This FR core twist probably initiated the slow rise motion to a height of torus stability domain as evidenced in the simulated field at 67th hour.

The eruptive behavior of the simulated magnetic field is further evaluated using several non-potential parameters. Over the three-day evolution, the relative magnetic helicity ($H_V$) steadily increases, reaching \(14\times10^{42}\,\text{Mx}^2\), while the free magnetic energy (\(E_f\)) rises to \(4\times10^{32}\) ergs, even as the total magnetic energy (\(E\)) slightly decreases due to the decay of the AR. The fractional free energy reaches 57\% for \(U = 150\,\text{m/s}\), indicating a stronger eruptive potential of the AR. The volume relative helicity is decomposed into a current-carrying component (\(H_J\)) and a mutual component between the current-carrying and potential fields (\(H_{PJ}\)), both of which increase steadily during the formation and further buildup of the flux rope. Notably, the ratio \(H_J/H_{V}\) attains a value of 0.13 at the time of flux rope formation ($\approx$50 hours) and exceeds 0.3 by the eruption time (67 hours).  These values are aligned with earlier observational and simulation results \citep{Pariat2017_HelRat, Zuccarello2018_Threshold_hj}, which suggest that while \(H_J/H_V \geq 0.1\) signifies an eruptive tendency, a ratio of \(H_J/H_V \geq 0.3\) is suggested to be linked with FR system that is prone to undergo torus instability and complete eruption. Indeed, the simulated FR core lies well above the torus-instability domain at the time of eruption, suggesting an association between torus instability and the threshold helicity ratio, though further case studies are needed to strengthen this relationship.

The data-driven MF simulations reproduce the magnetic field evolution of the AR, accurately capturing key observed features, including the timing of the flux rope formation and its topology. As these simulations are fully constrained by the observational data, the modeled evolution is considered to closely represent the actual solar conditions, making them a valuable tool for evaluating non-potential diagnostic parameters across different types of ARs.

{\bf Acknowledgments} SDO is a mission of NASA's Living With a Star Program. Field line rendering was done with the VAPOR visualization software (\url{www.vapor.ucar.edu/}). The MF simulations are carried out using the versatile, multi-user PENCIL code, which is publicly available at \url{https://pencil-code.nordita.org/}. H$\alpha$ images are obtained from Ha telescope at Kodaikanal Solar Observatory.  {\bf We are grateful to the anonymous reviewer for constructive comments that enhanced the objectivity of this work.}


\bibliographystyle{aasjournalv7}

\end{document}